\documentclass{PoS}

\usepackage[utf8x]{inputenc}

\usepackage{amssymb}
\usepackage{amsmath}

\def\P{{\boldsymbol P}}

\newcommand{\der}{\mathrm{d}}
\newcommand{\xt}{{{\boldsymbol x}_\perp}}
\newcommand{\yt}{{{\boldsymbol y}_\perp}}
\newcommand{\bt}{{{\boldsymbol b}_\perp}}
\newcommand{\rt}{{{\boldsymbol r}_\perp}}

\newcommand{\Pt}{{\P_\perp}}

\newcommand{\ud}{\, \mathrm{d}}
\newcommand{\tr}{\, \mathrm{Tr} \, }
\newcommand{\nc}{{N_\mathrm{c}}}

\newcommand{\Jpsi}{{J/\psi}}

\def\figscale{1.1}

\title{Forward $J/\psi$ production in pA collisions: centrality dependence}

\ShortTitle{Forward $J/\psi$ production in pA collisions: centrality dependence}

\author{\speaker{B. Duclou\'e}\\
	Department of Physics, P.O. Box 35, 40014 University of Jyväskylä, Finland\\
	and \\
	Helsinki Institute of Physics, P.O. Box 64, 00014 University of Helsinki,
	Finland \\
	E-mail: \email{bertrand.b.ducloue@jyu.fi}}

\author{T. Lappi\\
	Department of Physics, P.O. Box 35, 40014 University of Jyväskylä, Finland\\
	and \\
	Helsinki Institute of Physics, P.O. Box 64, 00014 University of Helsinki,
	Finland \\
	E-mail: \email{tuomas.v.v.lappi@jyu.fi}}

\author{H. Mäntysaari\\
	Physics Department, Brookhaven National Laboratory, Upton, NY 11973, USA\\
	E-mail: \email{mantysaari@bnl.gov}}

\abstract{The nuclear suppression of forward $J/\psi$ production in high energy proton-nucleus collisions can be used as a probe of gluon saturation at small $x$. In an earlier work we studied this suppression in minimum bias collisions in the Color Glass Condensate formalism, relying on the optical Glauber model to obtain the dipole cross section of the nucleus from the one of the proton fitted to HERA DIS data. Here we study how the impact parameter dependence of this model can be used to compare our results with recent LHC data on the centrality dependence of this suppression.}

\FullConference{XXIV International Workshop on Deep-Inelastic Scattering and Related Subjects\\
	11-15 April, 2016\\
	DESY Hamburg, Germany}

\begin{document}

\section{Introduction}

Forward $J/\psi$ meson production in high energy proton-nucleus collisions can provide valuable information about gluon saturation since it probes the target nucleus at very small $x$. This process can be studied in a perturbative approach due to the presence of a hard scale provided by the charm quark mass.
In a previous work~\cite{Ducloue:2015gfa} we studied this process in the Color Glass Condensate (CGC) formalism, using the Glauber approach to obtain the dipole cross section of a nucleus from the one of the proton. We showed that this improved nuclear geometry treatment leads to a smaller suppression in minimum bias collisions than in a previous study in this formalism~\cite{Fujii:2013gxa}, which is in better agreement with recent measurements of this observable at the LHC. More recently, the ALICE Collaboration measured the centrality dependence of the nuclear suppression in this process~\cite{Adam:2015jsa}. Here we investigate how the explicit impact parameter dependence of our model can be related to the centrality determination of such measurements

\section{Formalism}

In this work we use the simple Color Evaporation Model (CEM) to describe hadronization. In this model, it is assumed that a fixed fraction $F_{\Jpsi}$ of the $c\bar{c}$ pairs produced with an invariant mass lower than $2M_D$, where $M_D$ is the $D$-meson mass, will form $J/\psi$ bound states:
\begin{equation} 
\frac{\ud\sigma_{\Jpsi}}{\ud^2\P_{\perp}\ud Y}
=
F_{\Jpsi} \; \int_{4m_c^2}^{4M_D^2} \ud M^2
\frac{\ud\sigma_{c\bar c}}
{\ud^2\P_{\perp} \ud Y \ud M^2}
\, ,
\label{eq:dsigmajpsi}
\end{equation}
where $\P_{\perp}$ and $Y$ are the transverse momentum and rapidity of the produced $J/\psi$. The exact value of $F_{\Jpsi}$ is not important here since we will focus on ratios of cross sections in which it will simplify. 
The production of gluons and quark pairs in the dilute-dense limit of the Color Glass Condensate formalism was studied in Refs.~\cite{Blaizot:2004wu,Blaizot:2004wv} (see also Ref.~\cite{Kharzeev:2012py}) and used in a number of works, e.g. ~\cite{Fujii:2005rm,Fujii:2006ab,Fujii:2013gxa,Fujii:2013yja,Ma:2015sia}. This gives access to the $c\bar{c}$ pair production cross section in Eq.~(\ref{eq:dsigmajpsi}), which can be found in Ref.~\cite{Fujii:2013gxa}.

When $Y$, the rapidity of the produced $\Jpsi$, is large, the longitudinal momentum fraction $x_1$ probed in the projectile proton is large: $x_1=\sqrt{\Pt^2+M^2}e^{Y}/\sqrt{s}$. Therefore in this limit the gluon density in the proton can be described by a collinear PDF. The parametrization we use in practice is the LO MSTW 2008~\cite{Martin:2009iq} one. On the contrary, at large $Y$, the $x$ value probed in the target, $x_2$, is very small: $x_2=\sqrt{\Pt^2+M^2}e^{-Y}/\sqrt{s}$. Its gluon density is described by the dipole cross section $S_{_Y}(\xt-\yt) = \frac{1}{\nc }\left< \tr U^\dag(\xt)U(\yt)\right>$, 
where $U(\xt)$ is a Wilson line in the color field of the target in the fundamental representation. The rapidity evolution of $S_{_Y}(\rt)$ is governed by the Balitsky-Kovchegov (BK) equation~\cite{Balitsky:1995ub,Kovchegov:1999ua}. For a proton target we use dipole cross sections obtained by solving numerically the BK equation with running coupling corrections~\cite{Balitsky:2006wa} using the MV$^e$ parametrization~\cite{Lappi:2013zma} fitted to HERA DIS data~\cite{Aaron:2009aa} for the initial condition. Due to the lack of accurate nuclear DIS data it is not possible to perform a similar fit for a nuclear target. In this case we use, as in Ref.~\cite{Lappi:2013zma}, the Glauber approach to get the initial condition for a nucleus from the one of a proton. For this one assumes that the density of nucleons in the transverse plane follows the standard Woods-Saxon distribution $T_A(\bt)$, where $\bt$ is the impact parameter, and that at the initial rapidity of the BK evolution the high energy gluon coming from the projectile scatters independently off these nucleons. The nuclear density function $T_A(\bt)$ is the only additional input used to go from a proton to a nucleus. It also provides an explicit impact parameter dependence of the initial condition, from which one can solve the BK equation at each impact parameter.

\section{Results}

Since the impact parameter is not observable, experimental results are usually given in centrality classes instead. These are defined such that the $(0-c)\%$ most central collisions give $c\%$ of the total inelastic proton-nucleus cross section. To define a centrality class $(c_1-c_2)\%$ in the optical Glauber model, one would first compute the impact parameter values $b_1$ and $b_2$ corresponding to
\begin{equation}
	(c_1-c_2)\% = \frac{1}{\sigma_\text{inel}^{\text{pA}}} \int_{b_1}^{b_2} \der^2 \bt p(\bt),
	\label{eq:centrality}
\end{equation}
where the total inelastic proton-nucleus cross section $\sigma_\text{inel}^{\text{pA}}$ is given by $\sigma_\text{inel}^{\text{pA}} = \int \der^2 \bt \, p(\bt)$, with $p(\bt) = 1- e^{-A T_A(\bt) \sigma_\text{inel}^\text{pp}}$. The yield in each class is
\begin{equation}
	\frac{\der N}{\der^2 \Pt \der Y} = \frac{ \int_{b_1}^{b_2} \der^2 \bt \frac{\der N(\bt)}{\der^2 \Pt \der Y} } {\int_{b_1}^{b_2} \der^2 \bt \, p(\bt)},
\end{equation}
where $b_1$ and $b_2$ are calculated using Eq.~(\ref{eq:centrality}) and $\frac{\der N(\bt)}{\der^2 \Pt \der Y}$ is defined such that $\int \der^2 \bt \frac{\der N(\bt)}{\der^2 \Pt \der Y}=\frac{\der\sigma}{\der^2 \Pt \der Y}$.

\begin{table}[tb]
	\centering
	\begin{tabular}{|c|c|c|c|}
		\hline
		Centrality class & $\langle N_\text{coll} \rangle_\text{opt.}$ & $\langle N_\text{coll} \rangle_\text{ALICE}$
		&$\bt$ [fm]
		\\
		\hline
		2--10\%   & 14.7 & $11.7 \pm 1.2 \pm 0.9$ &  4.14\\
		10--20\%  & 13.6 & $11.0 \pm 0.4 \pm 0.9$ & 4.44 \\
		20--40\%  & 11.4 & $9.6 \pm 0.2 \pm 0.8$ & 4.94 \\
		40--60\%  & 7.7  & $7.1 \pm 0.3 \pm 0.6$ & 5.64 \\
		60--80\%  & 3.7  & $4.3 \pm 0.3 \pm 0.3$ & 6.29  \\
		80--100\% & 1.5  & $2.1 \pm 0.1 \pm 0.2$ & 6.91\\
		\hline
	\end{tabular}
	\caption{Average number of binary collisions $\langle N_\text{coll} \rangle$ in each centrality class in the optical Glauber model compared with the values estimated by ALICE~\cite{Adam:2015jsa}. The values of $\bt$ in the fourth column are obtained by solving $N_\text{bin}(\bt)=\langle N_\text{coll} \rangle_\text{ALICE}$.}
	\label{tab:Ncoll}
\end{table}

However, it is not possible to use this procedure to directly compare our results with the measurement of Ref.~\cite{Adam:2015jsa}. Indeed, if we compare the average number of binary nucleon-nucleon collisions in each class in the optical Glauber model,
\begin{equation}
\langle N_\text{coll} \rangle_\text{opt.} = \frac{\int_{b_1}^{b_2} \der^2 \bt N_\text{bin}(\bt)}{\int_{b_1}^{b_2} \der^2 \bt p(\bt)},
\end{equation}
where $N_\text{bin}(\bt) = A T_A(\bt) \sigma_\text{inel}^\text{pp}$, with the experimental estimate, we find a discrepancy between the two as can be seen from Table~\ref{tab:Ncoll}: for central collisions the $\langle N_\text{coll} \rangle$ values obtained in the optical Glauber model are larger than the experimental values, while they are smaller for peripheral collisions.
In the following, assuming that the $\langle N_\text{coll} \rangle$ values estimated by ALICE in each class are correct, we will base our comparison on the number of collisions instead of centrality classes.

\begin{figure}
	\centering
	\includegraphics[scale=\figscale]{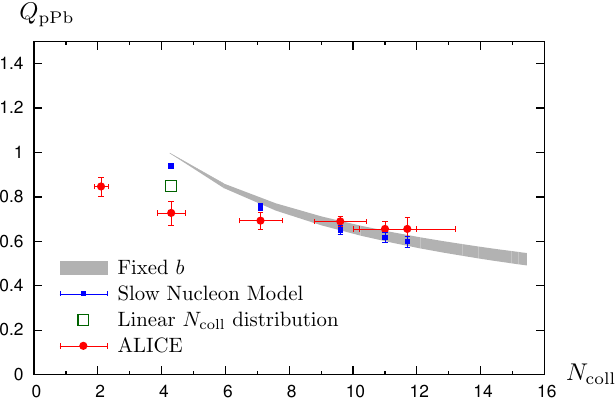}
	\hspace{0.3cm}
	\includegraphics[scale=\figscale]{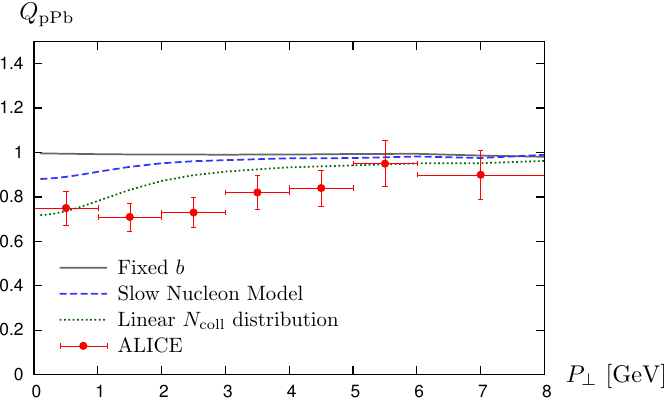}
	\caption{Nuclear modification factor $Q_\text{pPb}$ as a function of $N_\text{coll}$ (Left)	and as a function of $P_\perp$ in the $60-80\%$ centrality class (Right) at $\sqrt{s_{NN}}=5$ TeV. Data from Ref.~\cite{Adam:2015jsa}.}
	\label{fig:QpA_Ncoll}
\end{figure}

As a first approximation we will use a fixed impact parameter for each class, obtained by solving $N_\text{bin}(\bt)=\langle N_\text{coll} \rangle_\text{ALICE}$. The impact parameter values obtained in this way are shown in the fourth column of Table~\ref{tab:Ncoll}.
In Fig.~\ref{fig:QpA_Ncoll}~(L) we show the comparison of our results in this approach with ALICE data for $Q_{\rm pA}$, the nuclear modification factor defined as
\begin{equation}
Q_{\rm pA}= \frac{\frac{\ud N^\text{pA}}{\ud^2 \Pt \ud Y}}
{ A \langle T_A \rangle \frac{\ud\sigma^\text{pp}}{\ud^2 \Pt \ud Y}} \; ,
\end{equation}
as a function of $N_\text{coll}$. We observe that the agreement with data for central collisions is reasonable, but the variation with decreasing $N_\text{coll}$ is too strong. In particular, in our model we get $Q_{\rm pA} \sim 1$ at $N_\text{coll} \sim 4$ which is significantly above the data. In Fig~\ref{fig:QpA_Ncoll}~(R) we show $Q_{\rm pA}$ as a function of $\Pt$ in the $60-80\%$ centrality bin. Here we see that, contrary to the data, $Q_{\rm pA}$ is almost flat and very close to 1. This is due to the fact that in our model the saturation scale of the nucleus becomes as small as the one of the proton at an impact parameter $\bt \sim 6.3$ fm, as found in Ref.~\cite{Lappi:2013zma}, which corresponds to $N_\text{coll} \sim 4.3$~\cite{Ducloue:2016pqr}. This can be traced back to the fact that the value of the effective transverse area of the proton, $\sigma_0/2$, extracted from DIS fits is much smaller than the total inelastic proton-proton cross section~\cite{Lappi:2013zma}. This too strong dependence on centrality could probably be softened by taking a value of $\sigma_0/2$ of the order of $\sigma_\text{inel}^\text{pp}$, but this would break the consistent description of the proton from HERA to the LHC of our approach.

The fixed impact parameter approximation that we have used so far would be justified only if the fluctuations around $\langle N_\text{coll} \rangle$ are small in each centrality class considered by ALICE, which is not necessarily the case. For a more consistent comparison we will now consider distributions in the impact parameter space, obtained from two models of $N_\text{coll}$ distributions using the relation $N_\text{bin}(\bt) = A T_A(\bt) \sigma_\text{inel}^\text{pp}$.
First we use $N_\text{coll}$ distributions obtained in the Slow Nucleon Model (SNM)~\cite{Adam:2014qja} as provided by ALICE~\cite{alicecent} in each centrality class. The values of $\langle N_\text{coll} \rangle$ obtained in the SNM don't match the ones obtained with the hybrid method listed in Table~\ref{tab:Ncoll}, so we shift the distributions so that it is the case. We note that this method is biased~\cite{Adam:2014qja}, in contrast with the hybrid method used in Ref.~\cite{Adam:2015jsa} to obtain $\langle N_\text{coll} \rangle$ in each class. Preferably one would like to use $N_\text{coll}$ distributions obtained in an unbiased way but there is no such data for now.
To try to evaluate the importance of the exact shape of the $N_\text{coll}$ distributions, we also consider, in the $60-80\%$ centrality class (which is more sensitive to fluctuations than the more central ones), a simple distribution which decreases linearly with $N_\text{coll}$. This distribution depends on two parameters, the height at the origin and the $N_\text{coll}$ value where it reaches 0. These two parameters are fixed such that the distribution is normalized to 1 and $\langle N_\text{coll} \rangle=\langle N_\text{coll} \rangle_\text{ALICE}$. We emphasize that this linear distribution is not based on any detailed model but is used to illustrate the importance of the $N_\text{coll}$ distribution in this class.

The results obtained with these two models of $N_\text{coll}$ distributions are compared with ALICE data and the fixed impact parameter approximation in Fig.~\ref{fig:QpA_Ncoll}. We see that if we consider $Q_\text{pPb}$ as a function of $N_\text{coll}$, the values obtained with the SNM distribution are very close to the fixed $\bt$ approximation. On the other hand, the value obtained in the $60-80\%$ bin with the linear distribution is significantly closer to the data. Regarding $Q_\text{pPb}$ as a function of $\Pt$ in the $60-80\%$ bin, both models lead to values smaller than 1 at small $\Pt$. Here also the effect is stronger with the linear distribution than with the SNM one leading to results closer to the data.

\section{Conclusions}

In this work we extended our study of forward $\Jpsi$ suppression at the LHC to centrality dependent observables. We used, as in Ref.~\cite{Ducloue:2015gfa}, the optical Glauber model to extrapolate the proton dipole cross section to a nucleus. In this model the impact parameter dependence appears naturally.
However we have shown that it is not straightforward to relate it with the centrality determination at experiments.
Indeed, while our results for central collisions are in reasonable agreement with recent ALICE data, the model dependence for peripheral collisions seems to be quite large. For a more consistent comparison, one would therefore need to have access to an experimental determination of $N_\text{coll}$ distributions obtained in an unbiased way in each centrality class.

\section*{Acknowledgments}
T.~L. and B.~D. are supported by the Academy of Finland, projects
267321 and 273464. 
H.~M. is supported under DOE Contract No. DE-SC0012704.
This research used computing resources of 
CSC -- IT Center for Science in Espoo, Finland.
We would like to thank C. Hadjidakis and I. Lakomov for 
discussions on the ALICE data.
	
\providecommand{\href}[2]{#2}\begingroup\raggedright\endgroup
	
\end{document}